\documentclass[sigconf]{acmart}
\usepackage{graphicx}
\usepackage{float}
\usepackage{subfigure}
\usepackage{hyperref}
\usepackage[
    type={CC},
    modifier={by},
    version={4.0},
    imagewidth=0.38\linewidth
]{doclicense}
\usepackage{wrapfig}
\AtBeginDocument{%
  }

\copyrightyear{2023}
\acmYear{2023}
\setcopyright{rightsretained}
\acmConference[CHCHI 2023]{Chinese CHI 2023}{November 13--16, 2023}{Denpasar, Bali, Indonesia}
\acmBooktitle{Chinese CHI 2023 (CHCHI 2023), November 13--16, 2023, Denpasar, Bali, Indonesia}\acmDOI{10.1145/3629606.3629638}
\acmISBN{979-8-4007-1645-4/23/11}

\begin{document}

\title{Multi-channel Sensor Network Construction, Data Fusion and Challenges for Smart Home}

\author{He Zhang}
\orcid{0000-0002-8169-1653}
\email{hpz5211@psu.edu}
\affiliation{%
  \institution{The Future Laboratory, Tsinghua University}
  \streetaddress{No. 160 Chengfu Road}
  \city{Beijing}
  \country{P.R.C.}
  \postcode{100084}
}
\affiliation{%
  \institution{Pennsylvania State University}
  \streetaddress{Westgate Building}
  \city{University Park}
  \state{Pennsylvania}
  \country{USA}
  \postcode{16801}
}

\author{Robin Ananda}
\orcid{0000-0001-7126-1302}
\email{ylb18@mails.tsinghua.edu.cn}
\affiliation{%
  \institution{Academy of Arts \& Design, Tsinghua University}
  \streetaddress{Tsinghua University}
  \city{Beijing}
  \country{P.R.C.}
  \postcode{100084}
}

\author{Xinyi Fu}
\authornote{Corresponding author}
\orcid{0000-0001-6927-0111}
\email{fuxy@mail.tsinghua.edu.cn}
\affiliation{%
  \institution{The Future Laboratory, Tsinghua University}
  \streetaddress{No. 160 Chengfu Road}
  \city{Beijing}
  \country{P.R.C.}
  \postcode{100084}
}

\author{Zhe Sun}
\orcid{0000-0002-4821-2366}
\email{sunz15@tsinghua.org.cn}
\affiliation{%
  \institution{Academy of Arts \& Design, Tsinghua University}
  \streetaddress{Tsinghua University}
  \city{Beijing}
  \country{P.R.C.}
  \postcode{100084}
}
\author{Xiaoyu Wang}
\orcid{0009-0007-6812-7031}
\email{1800017830@pku.edu.cn}
\affiliation{%
  \institution{Yuanpei College, Peking University}
  \streetaddress{NO.5 Yiheyuan Road}
  \city{Beijing}
  \country{P.R.C}
  \postcode{100871}
}

\author{Keqi Chen}
\orcid{0000-0001-9383-9910}
\email{keqi001@e.ntu.edu.sg}
\affiliation{%
  \institution{Nanyang Technological University}
  \country{Singapore}
}

\author{John M. Carroll}
\orcid{0000-0001-5189-337X}
\email{jmc56@psu.edu}
\affiliation{%
  \institution{Pennsylvania State University}
  \streetaddress{Westgate Building}
  \city{University Park}
  \state{Pennsylvania}
  \country{USA}
  \postcode{16801}
}

\renewcommand{\shortauthors}{Zhang et al.}

\begin{abstract}
  Both sensor networks and data fusion are essential foundations for developing the smart home Internet of Things (IoT) and related fields. We proposed a multi-channel sensor network construction method involving hardware, acquisition, and synchronization in the smart home environment and a smart home data fusion method (SHDFM) for multi-modal data (position, gait, voice, pose, facial expression, temperature, and humidity) generated in the smart home environment to address the configuration of a multi-channel sensor network, improve the quality and efficiency of various human activities and environmental data collection, and reduce the difficulty of multi-modal data fusion in the smart home. SHDFM contains 5 levels, with inputs and outputs as criteria to provide recommendations for multi-modal data fusion strategies in the smart home. We built a real experimental environment using the proposed method in this paper. To validate our method, we created a real experimental environment — a physical setup in a home-like scenario where the multi-channel sensor network and data fusion techniques were deployed and evaluated. The acceptance and testing results show that the proposed construction and data fusion methods can be applied to the examples with high robustness, replicability, and scalability. Besides, we discuss how smart homes with multi-channel sensor networks can support digital twins.
\end{abstract}




\maketitle

\section{Introduction}
With the improvement of sensor technology, the development of artificial intelligence, the popularization of high-speed Internet, and the application of the concept of smart home Internet of Things, the concept of the smart home is being accepted by more and more people. At the same time, with the popularity of consumer-grade smart home products and the increase in human society's demand, the smart home market has a certain scale and is growing rapidly. Statista states that the global smart home market could reach \$182.45 billion by 2025~\cite{Statista19}. According to the scenario-based needs of users, a smart home mainly involves several key areas such as device control and linkage~\cite{zheng2017combining}, scene atmosphere~\cite{eggen2003exploring}, entertainment and leisure~\cite{kim2018advanced}, intelligent security~\cite{touqeer2021smart}, energy management~\cite{badar2022smart}, intelligent applications~\cite{kornyshova2022smart}, and design for specific people or needs~\cite{jamwal2022smart,schak2022smart, radha2022flexible}. On the one hand, there is a growing demand for smarter and more efficient smart homes and related products, and the various needs, motivations, goals, and implementation methods of using smart homes in smart home scenarios are repeatedly emphasized~\cite{marikyan2019systematic,hargreaves2018learning}. Among them, data and sensor networks are one of the most important foundations for studying smart home problems. On the other hand, traditional interaction methods have been found to have a number of interactivity problems that still affect the interaction experience, such as intrusiveness~\cite{benlian2020mitigating}, fatigue~\cite{liu2020real}, and interaction with specific groups of people~\cite{kim2021exploring}. The sensor network construction and data fusion application are the preconditions to help solve the above problems. In addition, humans have moved into the information society, where more and more interactions are moving from physical to virtual. Highly informative and virtualized interactions are gradually accepted and taken for granted. The diversity of data collected through multi-channel sensor networks significantly impacts future interaction methods, especially for further smart home digital twins.

There is currently no comprehensive summary of sensor network construction methods in the smart home. The main problems are the following two: (1) There are few comprehensive, integrated application and sensor network construction methods covering the whole smart home. Most of the existing studies are sensor networks built for specific studies, which are characterized by using only a small number of channels, a relatively single type, and a single purpose of sensors. Although similar sensor networks are still somewhat usable for specific research problems, with the further diversification of scenarios and requirements in smart homes, a series of unpredictable cross-influences on the research problems in real smart home environments cannot be effectively addressed or planned; (2) Existing studies rarely mention the specific implementation problems encountered in the construction of sensor networks, especially the configuration of hardware, acquisition, and synchronization in multi-channel sensor networks. The smart home is a significant research sub-field in human-computer interaction (HCI) and even in the interdisciplinary intersection. There are a lot of research problems related to people's livelihoods that can be explored~\cite{FUXy36}. At the same time, the a priori skills of researchers and technologies are likely to be limited in the interdisciplinary field, which limits further research. 

With the development of sensor technologies and the growing demand for HCI, ubiquitous computing~\cite{mehrotra2021overview} and data-driven~\cite{sarker2021data} approaches have received significant attention from researchers. The processing and fusion methods of the large amount of multimodal data generated in the smart home environment directly affect the effectiveness of the relevant models. The data fusion mentioned in this paper refers to the process of fusing real data from multiple sources, where disparate data are fused into usable integrated data. Traditional data fusion has certain challenges, mainly regarding uncertainty, integrity, and consistency after fusion [9]. In addition to the conventional challenges, data fusion solutions for smart homes also face the following new challenges:
\begin{enumerate}   
\item Existing data fusion solutions are not applicable to smart home problems.
\item Multimodal data fusion has limitations in terms of the types and numbers of modalities involved.
\item Only specific problems are addressed without global considerations, and portability is insufficient. 
\end{enumerate}  

Based on the above background and challenges, this paper firstly introduces the current situation, development trend, problems, and difficulties of multi-channel sensor network construction and data fusion and related research, and then proposes a multi-sensor network construction method applied to the smart home environment in conjunction with the development trend of smart home, future research-ability, sensor availability and related environment construction problems, and thus proposes a data fusion scheme in the smart home environment. Concurrently, the proposed methods for constructing the sensor network and the data fusion scheme were preliminarily tested through real scenario assessments on the comprehensive experimental platform we've built. 

\section{Related Work}

\subsection{The Construction of Multi-channel Sensor Networks}
The sensor network consists of multiple sensor nodes deployed in the smart home, forming a multi-hop self-organizing network using wireless communication. Liang et al.~\cite{liang2021smart} built a wireless sensor network for the smart home security system based on ZigBee protocol and other IoT technologies. Their proposed sensor network mainly consists of a digital access control system(hub) and sensor nodes. The sensor nodes mainly focus on detecting intrusion and indoor environmental information, including PIR, water level, smoke, and reed switch sensors. Ramson and Moni~\cite{ramson2017applications} reviewed various applications of wireless sensor networks, and the examples cited in their study regarding environmental monitoring, healthcare, and smart building categories are highly relevant to smart homes. Liu et al.~\cite{liu2023design} described the types of multi-channel sensors that may be used in smart environments and their uses.

Multi-device synchronization in multi-channel sensor networks poses significant challenges. While the widely used Network Time Protocol (NTP)~\cite{mills2010network} offers universal time coordination, it lacks precision (its time accuracy is low, providing 1-50ms time accuracy and relying on real-time network load), which may result in substantial errors in systems requiring high-frequency data, such as multi-camera video sensors. Additionally, to ensure the smooth functioning of the sensor network, a smart home system often requires a robust control base station, data acquisition, data processing, wireless communication, and power supply systems~\cite{shaikh2016energy}. While various studies~\cite{intille2005placelab,seo2021preference} have made valuable contributions to multi-channel sensor networks in smart homes, we aim to extend this research further. We aim to develop a system that integrates a broader array of channels with a more diverse range of sensors and smart home devices. Our proposed construction route is designed to be comprehensive, enhancing the system's overall quality. In our ongoing work, we strive for a system that excels in terms of wholeness, accessibility, robustness, reproducibility, and intelligence.

\subsection{Multi-channel Data Fusion}
Researchers have widely discussed data fusion and related techniques. The underlying logic is to combine different types, dimensions, and qualities of data from multiple sources and use them for any parameter estimation~\cite{castanedo2013review}. In related studies~\cite{ming2019data,sun2009information}, the conceptual terms of multimodal data fusion, multi-channel data fusion, and multisensor data fusion are often substitutable. Hall and Linas~\cite{hall1997introduction} proposed a definition of data fusion in which data from multiple sources are processed and combined based on relevant information to obtain higher accuracy and greater interpretability than single-source data. On the one hand, researchers have conducted a large number of studies related to multimodal data fusion, involving fusion strategies~\cite{chen2021new}, data processing methods~\cite{azcarate2021data}, multisensor data fusion~\cite{hall1997introduction}, information fusion~\cite{leung2019ai}, and data fusion applications~\cite{king2017application}, etc. Meanwhile, some technical engineering and methodological guidelines on data fusion are provided in related studies~\cite{liggins2017handbook,esteban2005review}. On the other hand, researchers have continuously reviewed and organized techniques and fusion methods related to data fusion~\cite{castanedo2013review,esteban2005review,durrant2016multisensor,khaleghi2013multisensor}.

The data fusion architecture system proposed by Dasarathy~\cite{dasarathy1997sensor} is one of the most famous data fusion methods, and its system consists of five categories based on input and output conditions, which are (1) Data In-Data Out (DAI-DAO): this method is the most basic data fusion method, where the raw data is fused, and the data itself is output directly after the sensor acquires the data; (2)Data In-Feature Out (DAI-FEO): Combine the sensors' original data and output the fused data's features; (3)Feature In-Feature Out (FEI-FEO): Both input and output are data features that are suitable for multi-source data with different data structures; (4) Feature In-Decision Out (FEI- DEO): Goal-oriented by inputting features and deriving decision labels based on a priori knowledge or pre-trained models. This category is often referred to as feature fusion; (5) Decision In-Decision Out (DEI- DEO): Both input and output are decisions, and this category is often referred to as decision fusion. All the five fusion strategies proposed in the method have some limitations. Among them, methods (1) to (4) are the common fusion methods, while method (5) has higher technical configuration requirements, especially for the deployment and configuration of full sensor networks. However, (5) has a good scope for development and is more suitable for the development trend in the context of artificial intelligence and machine learning. Based on this data fusion architecture, Fawzy et al.~\cite{fawzy2021spatiotemporal} proposed a spatiotemporal data fusion (STDF) method for low-level data input-output fusion for real-time spatial IoT resource aggregation by k-mean clustering for spatial clustering and tiny AG-gregation (TAG) for temporal aggregation, which reduced the data size by 95\% and saved 80\% of the processing time with 90\% accuracy.

Joint Directors of Laboratories (JDL)~\cite{white1991data}, although originating from the military domain, is currently the most popular data fusion model~\cite{khaleghi2013multisensor}, which is based on a data fusion process divided into five levels and includes an associated information database and an information channel for connecting the five levels, the database, the management system and the user interface. 

\subsection{Technology Mediation}
The use of information technology to create new "virtual spaces" is no longer just an idea but has become a happening today. Creations using virtual reality (VR)~\cite{eckstein2019smart}, augmented reality (AR)~\cite{ullah2012remote}, and mixed reality (MR)~\cite{speicher2019mixed} technologies are beginning to blur the boundaries of the world. Dourish's~\cite{dourish2006re} concept of relying on data and technology and developing new practices, new responses, and new environmental perceptions from everyday real-world experiences has been integrated into and become part of human society. The impact of data on the digital twin and the future of interaction is now widely recognized by academia. In the Smart Home domain, many possible devices and concept prototypes are made possible through data support. For example, the "Smart Home Virtual Tour" created by Ku\v{c}era and Haffner et al.~\cite{8337531} uses multiple sensors connected to an Arduino microcontroller to enable virtual tours and events to react to each other. Hu and Mao et al.~\cite{hu2022remote} proposed a digital twin and mixed reality based remote collaboration system for smart homes (RCSSH) in which cameras, environmental sensors (for temperature, harmful gas, smoke/fire), weight sensors, audio sensors (by headphones), and infrared sensors are deployed. Remote assistance is provided to users by collecting data from sensors in a mixed reality environment. Gopinath and Srija et al.~\cite{gopinath2019re} proposed a solution for redesigning the smart home using the digital twin, which relies on continuous connection and data transfer between sensors and virtual models and applications to enable real-time tracking and visualization of the smart home in a digital twin environment. However, there are few digital twins for the entire smart home environment.

\section{Multi-channel Sensor Network Construction for Smart Home}
\label{sec:Construction}
This paper proposes the whole sensor network system consists of four subsystems covering sensor acquisition, data processing, data transmission, database maintenance management, and automation functions, namely: (1) hardware system, (2) acquisition system, (3) synchronization system, and (4) robustness system. 

\subsection{Hardware, Data Collection and Synchronization System}
The hardware system of the smart home multi-channel sensor network contains (1) multi-channel sensors for collecting various activities and environmental data in the smart home environment; (2) Interprocess communication (IPC) system components for handling the data acquisition process; (3) Synchronization control system components for solving multi-channel sensor synchronization acquisition problems; (4) Network attached storage (NAS). Common types of sensors include multi-camera systems, microphone arrays, pressure sensors, electronic nose, gas sensors, sensors attached to furniture and home appliances, etc. The hardware system determines the upper limit of the quality of the collected data. It is recommended to focus on the sensor type, deployment location, synchronization method, and feasibility of the hardware system.

The multi-channel sensor network, as outlined in our real-time hardware model, has been deployed in a real-life smart home integrated experiment platform spread over approximately 60m$^{2}$. The setup comprises multiple FLIR cameras, connected to an Industrial PC (IPC) via a 10 GigE network, generating about 6.7TB of data daily, excluding bedrooms and bathrooms to maintain privacy. Microphones, olfactory, environmental, and specific device sensors, such as pressure and infrared, are positioned strategically. These sensors relay data to the IPC system using data cables, Bluetooth, or WiFi. The synchronized data is pre-processed and stored on a dedicated server for future use.

\section{Multi-channel Sensor Networks: Smart Home Data Fusion Model}
\begin{figure*}[!h]
\centering
  \includegraphics[width=\linewidth]{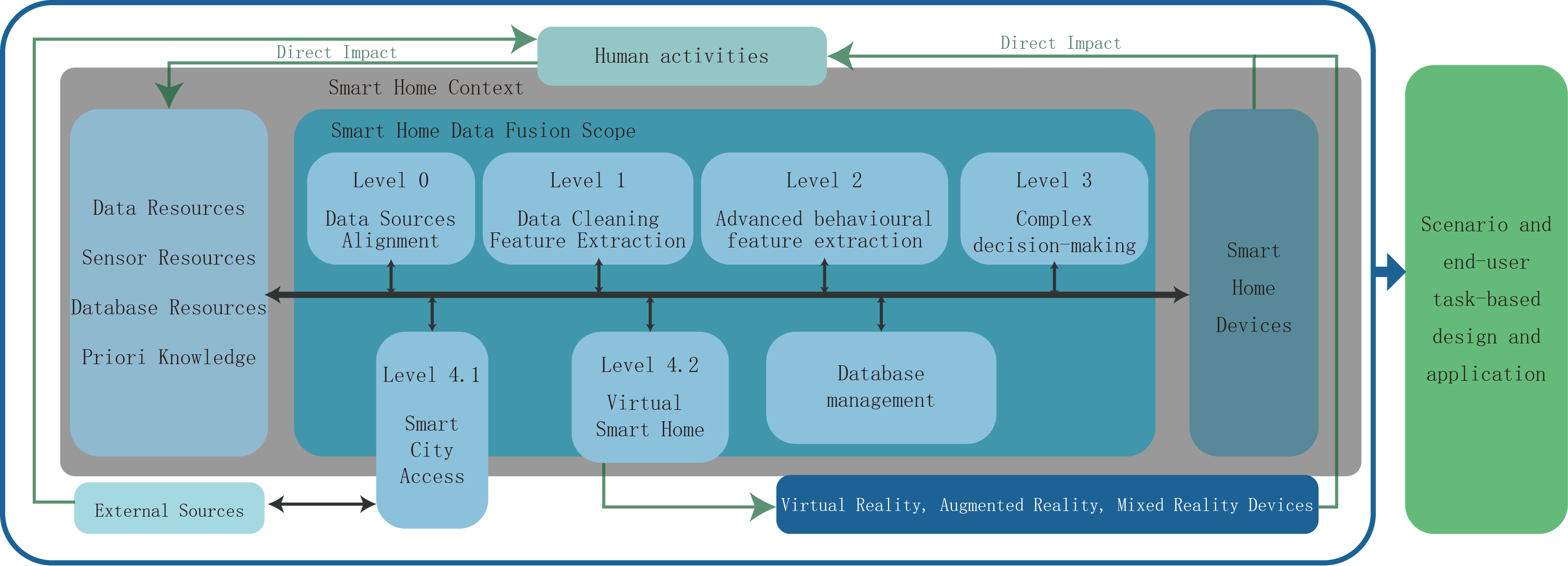}
  \caption{Smart Home Data Fusion Model (SHDFM): Human activity in the smart home environment directly influences the Resources of Data, Sensors, Database, and Prior knowledge on the left side of the diagram, which is processed through the information channel for smart home data fusion and then acts on smart home devices through data sources alignment, data cleaning feature extraction, advanced behavioral feature extraction, complex decision-making to influence human behavior. In addition, the data will also be used in smart cities through external resources after complex decisions have been made and through process optimization using virtual interactions to influence human behavioral activities. The final result is a Scenario and end-user task-based design and application.}
  \Description{SHDFM}
  \label{fig:SHDFM}
\end{figure*}

After the multi-channel sensor network system is built, human activity data and environmental data will be collected in the smart home environment. We propose a data fusion model for a multi-channel sensor network in smart home environments based on the JDL model in Figure~\ref{fig:SHDFM}, called Smart Home Data Fusion Model (SHDFM).
\begin{enumerate}   
\item[Level 0] This level implements the most basic data fusion, but unlike the traditional JDL model, the focus of this level is on multi-source data alignment, where the input is data from separate timelines and the output is data with the same timestamp. The synchronization components and timestamp alignment methods mentioned above are applied to align the multi-modal data in time units. At this level, the other modal data sizes can be aligned to the primary modal by specifying a few primary modalities. In the example, the amount of information contained in the images and audio is taken as a reference. The high-frequency data, such as the data of the multi-camera and microphone system, are chosen as the primary modality for alignment, considering the large amount of semantic information contained in the images and audio, but not in the sensors, such as environment and usage. 

This level addresses the critical limitation of mismatched timestamps often found in traditional data fusion methods. Here, rather than just fusing data, we lay emphasis on multi-source data alignment. The input comprises data from disparate timelines, and the output synchronizes them to have identical timestamps. By aligning high-frequency data, such as images and audio from multi-camera and microphone systems, which carry significant semantic information, to a primary modality, this level establishes a foundational solution for subsequent levels.

\item[Level 1] At this level, multi-model data sources are used for data filtering and labeling. For example, in a smart home environment using a multi-camera system, camera data is aligned at a specific frame rate. Main cameras are identified, unoccupied frames removed via person detection algorithms, and relevant frames are calibrated or deleted based on comparison with voice data. Building on the alignment achieved in Level 0, this level harnesses multi-modal data sources for data filtering and labeling. In practical scenarios like a smart home, this helps eliminate redundancy and enhances accuracy. 

Data from multi-channel sensors is fused to extract more useful data and features. Using the multi-camera system again as an example, a spatial coordinate system is established. Each camera's relative spatial coordinates are calibrated, and skeletal point coordinates are extracted and reconstructed in 3D, improving data accuracy. This, combined with aligned floor data, helps obtain the subject's movements and position in the room.

Ideally, the processes of these initial stages can feed into and optimize the outputs of further stages.

\item[Level 2] This level involves further extraction of data features and high-level states, such as types of activities, emotions, health conditions, and interactions between objects. Here, the labeling results are more complex, influenced by prior knowledge and human factors. Also, the fusion process goes through cycles for optimization, knowledge updates, and expansion of dynamic variable influence on the output. Studies at this level focus on advanced state behavior recognition, multi-modal emotion detection, and specific interaction studies. For instance, in ongoing research, researchers are trying to identify thirteen advanced states of occupants based on camera and floor sensor data (see example in the Figure~\ref{fig:Behavior-posture}). Furthermore, they performed emotion recognition based on gait and speech information, categorizing the multi-modal data into nine emotional states.

\item[Level 3] Complex Decision Making: The advanced tags, such as various human activities, emotions, events, and behaviors in the smart home environment obtained in the previous level, are fused and combined with environmental data for complex decision-making or higher-level activity recognition. Compared to level 2, level 3 has more uncertainty in the output but more personalization. Also, from this level onwards, there will be interactions initiated by things to people, such as active recommendation systems and emotional interventions. Therefore, from this level, intelligent interactive devices within the smart home system are invoked.

\item[Level 4.1] Smart City Access: External resources beyond the limits of the smart home begin to be called upon, such as big data and more shared data beyond the scope of the smart home. This level can result from quantitative changes, such as independent smart home networks accessing the broader smart city. At the core of this is the fact that sensor channels and coverage areas reach a sizable scale in a given area of human activity, and data sharing is no longer bound by space and time
, and the data can support multiple events and context awareness.

\item[Level 4.2] Virtual Smart Home: it represents an untapped yet promising avenue parallel to level 4.1. It encompasses the virtualization of a smart home, achieved through continuous data feature interaction, extraction, recognition, fusion, and decision-making. The integral processes of this level and level 4.1 mutually influence each other without a hierarchical relationship. The multi-channel sensor network data can be rendered into a virtual space through digital twins, establishing a connection between actions within the virtual space and the real world. This virtual smart home transcends traditional notions of space and time, entirely supported by the underlying data infrastructure.


\end{enumerate}  

SHDFM weaves the smart home environment with a multi-channel sensor network in a dynamic, adaptable architecture. As we climb the SHDFM ladder, requirements and resources evolve, becoming more intricate. Interestingly, as the complexity increases, the volume of output data decreases, a testament to the refined processes at play. While each level has the flexibility for self-iteration and influences across levels, they all present vast research potential.

Presently, extensive research efforts are directed toward levels 0 to 3, and with technological advancements, level 4 is also being explored. Moving forward, we anticipate continued improvements across levels 0 to 3, potentially in a non-linear fashion. Moreover, we foresee the exploration of levels 4.1 and 4.2 as the focus of future work, provided the resource ceiling is lifted. However, delving into these advanced levels may bring ethical considerations into sharper focus.

\section{Further Discussion}
The smart home environment introduces unique challenges that conventional data fusion methods often need to be equipped to tackle. Traditional solutions need to improve in the face of smart homes' dynamic nature, diversity of data, and the pivotal role of privacy and security.

Smart homes are characterized by the simultaneous evolution of numerous factors, including residents' behaviors, varied ambient conditions, and the status of interconnected smart devices. These homes generate vast data types, requiring a flexible fusion strategy to manage disparate data scales efficiently. Additionally, a high emphasis on privacy and security becomes necessary, which isn't typically a primary concern in other data fusion scenarios.

SHDFM is designed to address these specific challenges, utilizing a non-unidirectional flow architecture that permits dynamic adjustments and iterative optimization across various layers. This novel approach facilitates the model's adept adaptation to the dynamic nature of smart homes.

Conventional data fusion methods, founded on a linear or unidirectional data processing flow, need help to adapt efficiently to smart homes' dynamism and diverse data types. Furthermore, they may need more mechanisms to ensure adequate privacy and security protections. While these methods could be modified to a degree, they may not offer robust or optimal solutions in the smart home context.

Our proposed SHDFM, with its structured yet adaptable approach, confronts these challenges head-on while vigilantly maintaining user privacy and security. As we continue to refine this model, we anticipate it is playing a transformative role in advancing the capabilities and user-friendliness of future smart home systems.

\section{Future Work}
One of our ongoing works focuses on advancing the application of Digital Twin technology in our smart home system. We are striving to optimize the multi-modal data collection from our network of over 100 IoT devices and enhance the accuracy of our smart home virtual representation. In parallel, we enrich our understanding of human-centric data through our multi-channel sensor network, emphasizing capturing and interpreting behavioral and cognitive insights. The ultimate goal is to extend our capabilities to virtually map human interactions within the smart home, providing a comprehensive picture of user behavior.

The prototype (Figure~\ref{fig.Digital Twin}) we've created is a detailed replication of a real-world smart home environment, capturing key elements such as furniture, devices, and sensor placements. Its objective is to provide an extensive visual interpretation of the smart home setup. Our next step involves upgrading the prototype to allow interaction with specific devices within this virtual space, driven by data from the actual smart home. This enhancement would be possible through further improvements in the prototype's capabilities.   

Additionally, we are progressing with a project that leverages the SHDFM to analyze and synthesize primary behavioral data along with environmental data. The goal of this project is to extract high-level semantic information like human emotion, intent, and cognition~\cite{zhang2023decoding}. For a more thorough digital twin, we gather human data through multi-channel sensors, which aid in affective analysis.

\section{Conclusion}
This paper delves into the research landscape of smart homes, scrutinizes multi-channel sensor networks and data fusion literature specific to this realm, and pinpoints existing challenges. We propose a robust method for constructing multi-channel sensor networks and present a comprehensive example as a reference for researchers, particularly those venturing into interdisciplinary fields. We further introduce a modified version of the JDL data fusion model, dubbed the SHDFM, to establish a blueprint for smart home data fusion studies. To date, we have accomplished Level 0 and 1, along with aspects of Level 3 and 4.2, with the remaining levels slated for future work.

Key areas of focus for next-generation smart home sensor networks include amplifying network transmission rates and incorporating snap-in sensors that are simpler to install. The emphasis on Level 4 of the SHDFM is apparent and poised to grow. Overcoming technical obstacles while addressing ethical concerns affiliated with smart homes remains paramount. Furthering our understanding of user psychology and needs will allow for more targeted and practical frameworks and platforms.

Our proposed framework and methodology are designed to aid in constructing smart home research platforms, underpin theoretical research on sensor networks and data fusion, and guide the development of smart home products. By providing examples of smart home platforms built on the foundation of this framework and theory, we aim to streamline subsequent research. Looking forward, the creation of advanced smart home experimental platforms, incorporating comprehensive sensor networks and data fusion, will stand as a crucial focus for our future endeavors in HCI studies.

\begin{acks}
This work was supported by the National Natural Science Foundation of China (Grant No. 62172252)
\end{acks}

\bibliographystyle{ACM-Reference-Format}
\bibliography{sample-base}

\appendix
\section{APPENDICESx}

\begin{figure*}[htbp]
    \centering
    \subfigure[Example of behavior-posture annotation. The pose estimation diagram with background is shown in the upper left corner. The upper right corner shows the pose estimation without background. The lower left corner shows the gesture estimation with background, including hand gestures. The lower right corner shows the gesture estimation with no background and hand gestures.]{
    \label{fig:Behavior-posture}
    \includegraphics[width=0.8\linewidth]{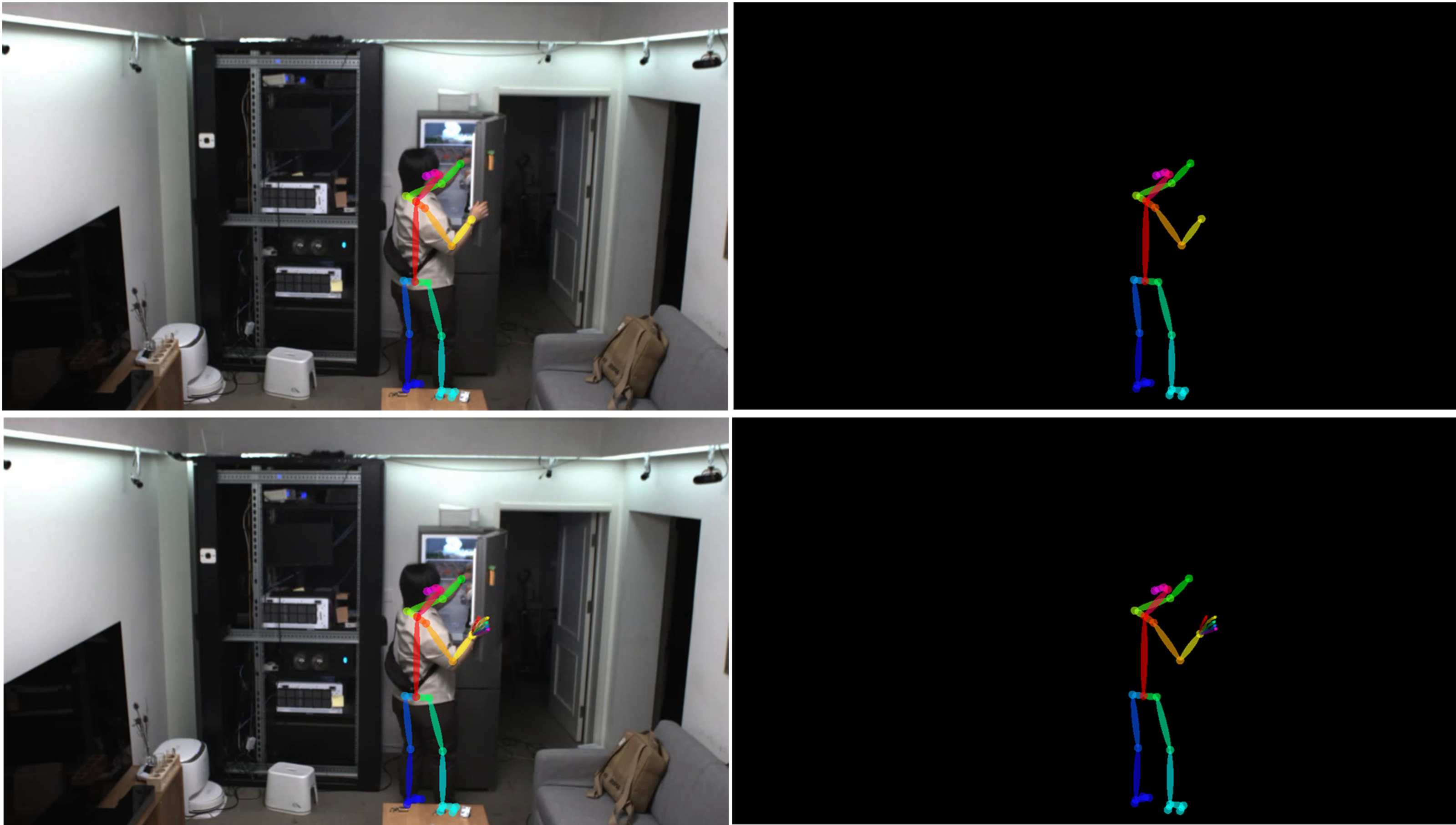}
    }
    \subfigure[Smart Home Digital Twin Prototype. The prototype shows an overview of the entire smart home environment and the deployment status of the multi-channel sensor. The diagram includes the house space content, furniture, devices, and sensors deployment.]{
    \label{fig.Digital Twin}
    \includegraphics[width=0.8\linewidth]{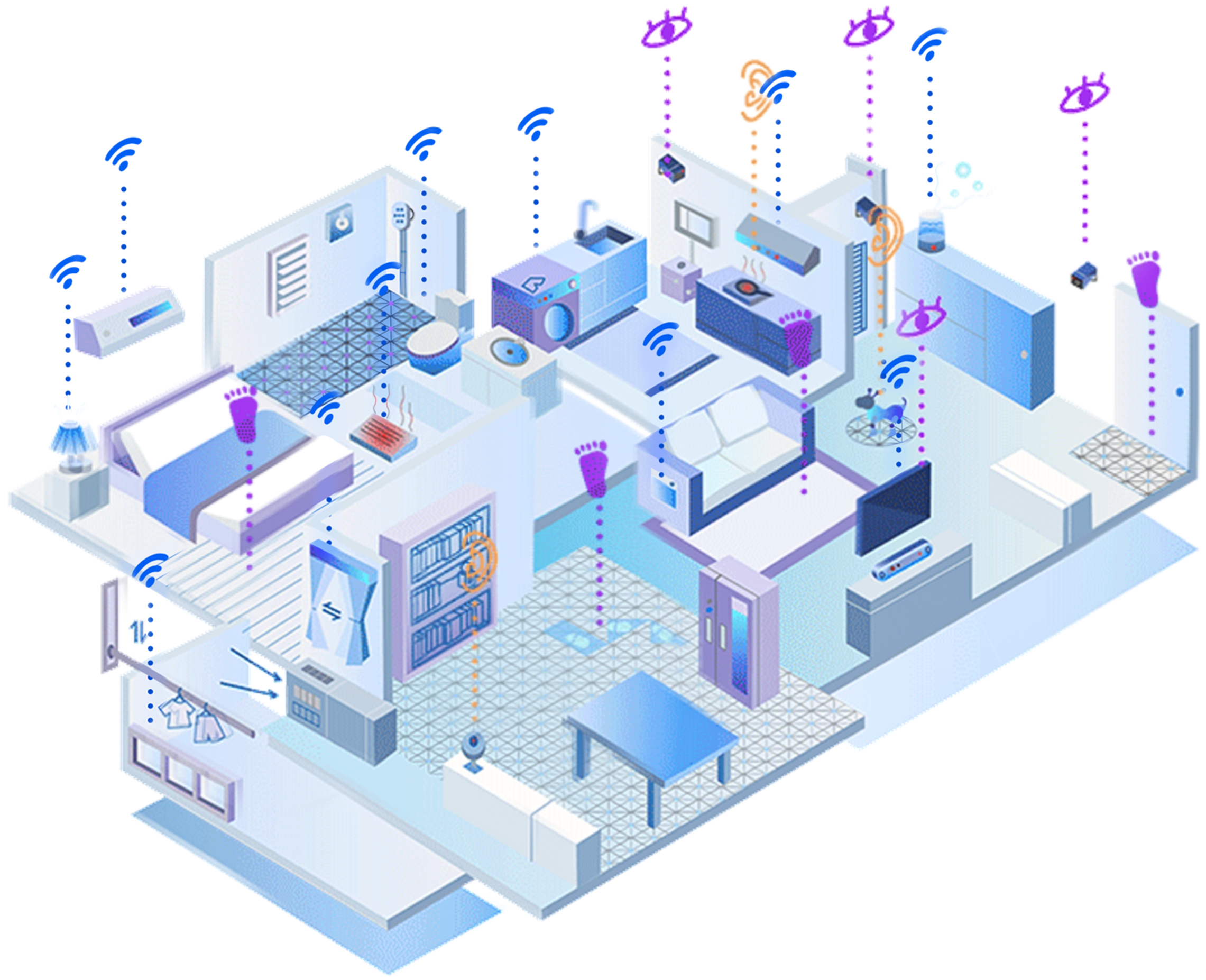}
    }
    \caption{Example of Behavior-posture Annotation and Smart Home Digital Twin Prototype.}
    \label{fig.Example}
\end{figure*}

\end{document}